\documentclass[journal]{IEEEtran}
\usepackage{hyperref}
\usepackage{graphicx}
\usepackage{color}
\usepackage{multirow}
\begin{document}

\title{Edge Intelligence: The Confluence of\\Edge Computing and Artificial Intelligence}
\author{
  Shuiguang~Deng,~\IEEEmembership{Senior~Member,~IEEE},
  Hailiang~Zhao,
  Weijia~Fang,
  Jianwei~Yin,
  Schahram~Dustdar,~\IEEEmembership{Fellow,~IEEE},
  and~Albert~Y.~Zomaya,~\IEEEmembership{Fellow,~IEEE}
  \thanks{This research was partially supported by the National Key Research and Development Program of China 
  (No.2017YFB1400601), National Science Foundation of China (No.61772461 \& No.61825205) and Natural Science 
  Foundation of Zhejiang Province (No.LR18F020003).}
  \thanks{S. Deng is with the First Affiliated Hospital, Zhejiang University School of Medicine, 
    310003 Hangzhou China and also the College of Computer Science and Technology, 
    Zhejiang University, 310058 Hangzhou, China. E-mail: dengsg@zju.edu.cn.}
  \thanks{H. Zhao and J. Yin are with the College of Computer Science and Technology, 
    Zhejiang University, 310058 Hangzhou, China. E-mail: \{hliangzhao, zjuyjw\}@zju.edu.cn.}
  \thanks{W. Fang is with the First Affiliated Hospital, Zhejiang University School of Medicine, 
  310003 Hangzhou, China. E-mail: weijiafang@zju.edu.cn,}
  \thanks{S. Dustdar is with the Distributed Systems Group, Technische Universität Wien, 1040 
  Vienna, Austria. E-mail: dustdar@dsg.tuwien.ac.at.}
  \thanks{A. Y. Zomaya is with the School of Computer Science, University of Sydney, Sydney, 
  NSW 2006, Australia. E-mail: albert.zomaya@sydney.edu.au.}
  \thanks{Weijia Fang is the corresponding author.}
}


\IEEEtitleabstractindextext{%
\begin{abstract}
Along with the rapid developments in communication technologies and the surge in the use of mobile devices, 
a brand-new computation paradigm, Edge Computing, is surging in popularity. Meanwhile, Artificial 
Intelligence (AI) applications are thriving with the breakthroughs in deep learning and the many improvements 
in hardware architectures. Billions of data bytes, generated at the network edge, put massive demands 
on data processing and structural optimization. Thus, there exists a strong demand to integrate 
Edge Computing and AI, which gives birth to Edge Intelligence. In this paper, we divide Edge 
Intelligence into AI for edge (Intelligence-enabled Edge Computing) and AI on edge (Artificial 
Intelligence on Edge). The former focuses on providing more optimal solutions to key problems 
in Edge Computing with the help of popular and effective AI technologies while the latter studies 
how to carry out the entire process of building AI models, i.e., model training and inference, 
on the edge. This paper provides insights into this new inter-disciplinary field from a 
broader perspective. It discusses the core concepts and the research road-map, which 
should provide the necessary background for potential future research initiatives in Edge Intelligence. 
\end{abstract}

\begin{IEEEkeywords}
  Edge Intelligence, Edge Computing, Wireless Networking, Computation Offloading, Federated Learning.
\end{IEEEkeywords}}

\maketitle

\IEEEdisplaynontitleabstractindextext

\ifCLASSOPTIONpeerreview
\begin{center} \bfseries EDICS Category: 3-BBND \end{center}
\fi
%
\IEEEpeerreviewmaketitle

\section{Introduction} \label{Sec1}
\IEEEPARstart{C}{ommunication} technologies are undergoing a new revolution. The advent of the 
5th generation cellular wireless systems (5G) that brings enhanced Mobile BroadBand (eMBB), Ultra-Reliable 
Low Latency Communications (URLLC) and massive Machine Type Communications (mMTC). With the proliferation 
of the Internet of Things (IoTs), more data is created by widespread and geographically distributed 
mobile and IoT devices, and probably more than the data generated by the mega-scale cloud datacenters \cite{IoT2}. Specifically, according 
to the prediction by Ericsson, 45\% of the 40ZB global internet data will be generated by IoT devices in 
2024 \cite{ericsson}. Offloading such huge data from the edge to cloud is intractable because it can lead 
to excessive network congestion. Therefore, a more applicable way is to handle user demands at the edge 
directly, which leads to the birth of a brand-new computation paradigm, (Mobile $\to$ Multi-access) Edge 
Computing \cite{IoT3}. The subject of Edge Computing spans many concepts and technologies in diverse 
disciplines, including Service-oriented Computing (SOC), Software-defined Networking (SDN), Computer 
Architecture, to name a few. The principle of Edge Computing is to push the computation and communication resources 
from cloud to edge of networks to provide services and perform computations, avoiding unnecessary 
communication latency and enabling faster responses for end users. Edge Computing is a booming field today. 

No one can deny that Artificial Intelligence (AI) is developing rapidly nowadays. Big data processing necessitates 
that more powerful methods, i.e., AI technologies, for extracting insights that lead to better decisions and strategic 
business moves. In the last decade, with the huge success of AlexNet and Deep Neural Networks (DNNs), which can learn the 
deep representation of data, have become the most popular machine learning architectures. Deep learning, represented by 
DNNs and their offshoots, i.e., Convolutional Neural Networks (CNNs), Recurrent Neural Networks (RNNs) and Generative 
Adversarial Networks (GANs), has gradually become the most popular AI methods in the last few years. Deep learning has 
made striking breakthroughs in a wide spectrum of fields, including computer vision, speech recognition, natural language 
processing, and board games. Besides, hardware architectures and platforms keep on improving at a rapid rate, which makes 
it possible to satisfy the requirements of the computation-intensive deep learning models. Application-specific accelerators 
are designed for further improvement in throughput and energy efficiency. In conclusion, driven by the breakthroughs in deep 
learning and the upgrade of hardware architectures, AI is undergoing sustained success and development. 

\begin{table*}[!ht]
  \renewcommand{\arraystretch}{1.2}
  \caption{Related Surveys and Their Emphases.}
  \label{table1}
  \centering
  \begin{tabular}{ccc}
    \hline\hline
    \textbf{Perspectives} & \textbf{Related Surveys} & \textbf{Highlights} \\
    \hline
      Intelligent Wireless Networking & 
      \cite{review-add1} \cite{review-add2} \cite{wireless_survey} & 
      \multicolumn{1}{l}{
        \begin{minipage}{3.5in}
          \vskip 4pt
          \begin{itemize}
            \item Summarize the utilization of machine learning on the wireless edge
            \item Including basic principles and general applications
            \item Focus on resource management, networking, and mobility management
            \item Optimization across different layers with machine learning technologies
          \end{itemize}
          \vskip 4pt
        \end{minipage}
      } \\
      \hline
      Definitions and Divisions of Edge Intelligence & 
      \cite{survey} \cite{openEI} \cite{review-add3} & 
      \multicolumn{1}{l}{
        \begin{minipage}{3.5in}
          \vskip 4pt
          \begin{itemize}
            \item Motivation, definition, division of Edge Intelligence
            \item Including architectures, enabling technologies, learning frameworks, and software platforms 
            \item Focus on model training and inference on edge
            \item Discuss the application scenarios and the practical implementations
          \end{itemize}
          \vskip 4pt
        \end{minipage}
      } \\
    \hline
    \hline
  \end{tabular}
\end{table*}

Considering that AI is functionally necessary for the quick analysis of huge volumes of data and extracting insights, there 
exists a strong demand to \textit{integrate} Edge Computing and AI, which gives rise to Edge Intelligence. Edge 
Intelligence is not the simple combination of Edge Computing and AI. The subject of Edge Intelligence is tremendous and 
enormously sophisticated, covering many concepts and technologies, which are interwoven together in a complex manner. 
Some works study the concept from the perspective of constructing \textit{Intelligent Wireless Networks on Edge}. 
For example, Sun et al. comprehensively survey the recent advances of the applications of machine learning technologies 
in wireless communication \cite{review-add1}. Specifically, this paper classifies the utilization of machine learning on 
the wireless edge into three parts: resource management in the MAC layer, networking and mobility management in the network 
layer, and localization in the application layer. Different conditions and unsolved challenges are also discussed. In addition, 
Mao et al. study the state-of-the-art researches on the applications of deep learning algorithms for different network layers 
\cite{review-add2}. Currently, there is no formal and internationally accepted definition of Edge Intelligence. To 
deal with the problem, some researchers put forward their definitions. For example, Zhou et al. believe that the scope of Edge 
Intelligence should not be restricted to running AI models solely on edge servers or devices but in collaboration of edge and cloud \cite{survey}. They define six levels of Edge Intelligence, from \textit{cloud-edge co-inference} 
(level 1) to \textit{all on-device} (level 6). Zhang et al. define Edge Intelligence as the capability to enable edges to 
execute AI algorithms \cite{openEI}. In Table \ref{table1}, we summarize related survey papers on edge intelligence. 

In this paper, we propose to establish a \textit{broader} vision and perspective. We suggest to distinguish edge Intelligence 
into \textit{AI for edge} and \textit{AI on edge}. 
\begin{enumerate}
  \item \textit{AI for edge} is a research direction focusing on providing a better solution to constrained 
  optimization problems in Edge Computing with the help of effective AI technologies. Here, AI is used to 
  endow edge with more intelligence and optimality. Therefore, it can be understood as \textit{Intelligence-enabled 
  Edge Computing (IEC)}. 
  \item \textit{AI on edge} studies how to run AI models on edge. It is a framework for 
  running training and inference of AI models with device-edge-cloud synergy, which aims at extracting insights 
  from massive and distributed edge data with the satisfaction of algorithm performance, cost, privacy, reliability, 
  efficiency, etc. Therefore, it can be interpreted as \textit{Artificial Intelligence on Edge (AIE)}. 
\end{enumerate}
Edge Intelligence, currently in its early stage, and is attracting more researchers and companies from all over the 
world. To disseminate the recent advances of Edge Intelligence, Zhou et al. have conducted a comprehensive and concrete 
survey of the recent research efforts on Edge Intelligence \cite{survey}. They survey the architectures, enabling technologies, 
systems, and frameworks from the perspective of AI models' training and inference. However, the material in Edge Intelligence 
spans an immense and diverse spectrum of literature, in origin and in nature, which is not fully covered by this survey. Many 
concepts are still unclear and questions remain unsolved. The research process actually what motivated us to write this paper to 
shed some light and provide more insights \textit{with simple and clear classification}. 

We commit ourselves to elucidating Edge Intelligence to provide a  broader vision and perspective. In Section \ref{Sec2}, we 
discuss the relation between Edge Computing and AI. In Section \ref{Sec3}, we demonstrate the research road-map of 
Edge Intelligence concisely with a hierarchical structure. Section \ref{Sec4} and Section \ref{Sec5} elaborate the 
state of the art and grand challenges on \textit{AI for edge} and \textit{AI on edge}, respectively. Section \ref{Sec6} 
concludes the article.

\section{The Relations between Edge Computing and AI} \label{Sec2}
We believe that the confluence of AI and Edge Computing is natural and inevitable. In effect, there is an interactive 
relationship between them. On one hand, AI provides Edge Computing with technologies and methods, and Edge Computing 
can unleash its potential and scalability with AI; on the other hand, Edge Computing provides AI with scenarios and 
platforms, and AI can expand its applicability with Edge Computing. 

\textbf{AI provides Edge Computing with technologies and methods.} In general, Edge Computing is a distributed 
computing paradigm, where software-defined networks are built to decentralize data and provide services with 
robustness and elasticity. Edge Computing faces resource allocation problems in different layers, such as CPU cycle 
frequency, access jurisdiction, radio-frequency, bandwidth, and so on. As a result, it has great demands on various 
powerful optimization tools to enhance system efficiency. AI technologies are capable to handle this task. Essentially, 
AI models extract unconstrained optimization problems from real scenarios and then find the asymptotically optimal 
solutions iteratively with Stochastic Gradient Descent (SGD) methods. Either statistical learning methods or deep 
learning methods can offer help and advice for the edge. Besides, reinforcement learning, including multi-armed bandit 
theory, multi-agent learning and deep $Q$-network (DQN), is playing a growing and  important role in resource allocation 
problems for the edge.

\textbf{Edge Computing provides AI with scenarios and platforms.} The surge of IoT devices makes the Internet of 
Everything (IoE) a reality \cite{IoT5}. More and more data is created by widespread and geographically distributed 
mobile and IoT devices, other than the mega-scale cloud datacenters. Many more application scenarios, such as 
intelligent networked vehicles, autonomous driving, smart home, smart city and real-time data processing in public 
security, can greatly facilitate the realization of AI from theory to practice. Besides, AI applications with high 
communication quality and low computational power requirements can be migrated from cloud to edge. In a word, Edge 
Computing provides AI with a heterogeneous platform full of rich capabilities. Nowadays, it is gradually becoming possible 
that AI chips with computational acceleration such as Field Programmable Gate Arrays (FPGAs), Graphics Processing Units 
(GPUs), Tensor Processing Units (TPUs) and Neural Processing Units (NPUs) are integrated with intelligent mobile 
devices. More corporations participate in the design of chip architectures to support the edge computation 
paradigm and facilitate DNN acceleration on resource-limited IoT devices. The hardware upgrade on edge also injects 
vigor and vitality into AI.

\section{Research road-map of Edge Intelligence} \label{Sec3}

\begin{figure*}[htbp]
  \centering
  \includegraphics[width=5.3in]{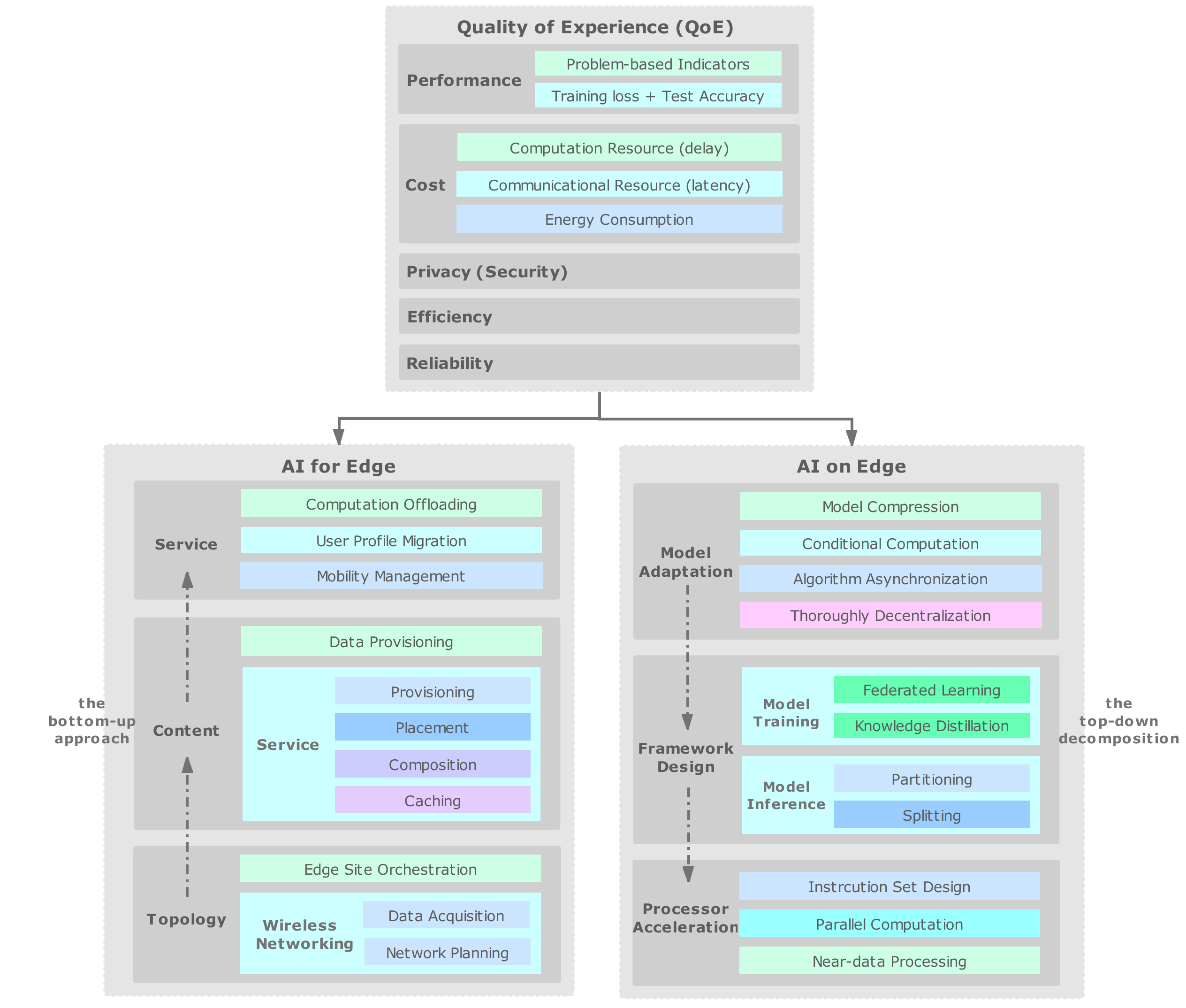}
  \caption{The research road-map of Edge Intelligence.}
  \label{fig1}
\end{figure*}

The architectural layers in the Edge Intelligence road-map, depicted in Fig. \ref{fig1}, describe a logical separation 
for the two directions respectively, i.e., \textit{AI for edge} (left) and \textit{AI on edge} (right). In the bottom-up 
approach, we divide research efforts in Edge Computing into Topology, Content, and Service. AI technologies can be utilized 
in all of them. By top-down decomposition, we divide the research efforts in \textit{AI on edge} into Model Adaptation, 
Framework Design and Processor Acceleration. Before discussing \textit{AI for edge} and \textit{AI on edge} separately, 
we first describe the goal to be optimized for both of them, which is collectively known as Quality of Experience (QoE). 
QoE remains at the top of the road-map.

\subsection{Quality of Experience} \label{Sce3.1}
We believe that QoE should be \textit{application-dependent} and determined by jointly considering multi-criteria: Performance, 
Cost, Privacy (Security), Efficiency and Reliability. 

1) \textbf{Performance}. Ingredients of \textit{performance} are different for \textit{AI for edge} and \textit{AI on edge}. 
As for the former, performance indicators are problem-dependent. For example, performance could be \textit{the ratio of 
successfully offloading} when it comes into the computation offloading problems. It could be the service providers' 
\textit{need-to-be-maximized revenue} and \textit{need-to-be-minimized hiring costs} of Base Stations (BSs) when it comes 
into the service placement problems. As for the latter, performance mainly consists of training loss and inference 
accuracy, which are the most important criteria for AI models. Although the computation scenarios have changed from cloud 
clusters to the synergised system of device, edge, and cloud, these criteria still play important roles.

2) \textbf{Cost}. Cost usually consists of computation cost, communication cost, and energy consumption. Computation cost 
reflects the demand for computing resources such as achieved CPU cycle frequency, allocated CPU time while communication 
cost presents the request for communication resources such as power, frequency band and access time. Many works also focused 
on minimizing the delay (latency) caused by allocated computation and communication resources. Energy consumption is not 
unique to Edge Computing but more crucial due to the limited battery capacity of mobile devices. Cost reduction is crucial 
because Edge Computing promises a dramatic reduction in delay and energy consumption by tackling the key challenges for 
realizing 5G.

3) \textbf{Privacy (Security)}. With the increased awareness of the leaks of public data, privacy preservation has become 
one of the hottest topics in recent years. The status quo led to the birth of Federated Learning, which aggregates local 
machine learning models from distributed devices while preventing data leakage \cite{federated1}. The security is closely 
tied with privacy preservation. It also has an association with the robustness of middleware and software of edge systems, 
which are not considered in this article.

4) \textbf{Efficiency}. Whatever \textit{AI for edge} or \textit{AI on edge}, high efficiency promises us a system with 
excellent performance and low overhead. The pursuit of efficiency is the key factor for improving existing algorithms and 
models, especially for \textit{AI on edge}. Many approaches such as model compression, conditional computation, and algorithm 
asynchronization are proposed to improve the efficiency of training and inference of deep AI models.

5) \textbf{Reliability}. System reliability ensures that Edge Computing will not fail throughout any prescribed operating 
periods. It is an important indicator of user experience. For Edge Intelligence, system reliability appears to be particularly 
important for \textit{AI on edge} because the model training and inference are usually carried out in a distributed and 
synchronized way and the participated local users have a significant probability of failing to complete the model upload and 
download due to wireless network congestion.

\subsection{A Recapitulation of IEC} \label{Sce3.2}
The left-side of the road-map, depicted in Fig. \ref{fig1}, is \textit{AI for edge}. We name this kind of work IEC (i.e. 
Intelligence-enabled Edge Computing) as AI provides powerful tools for solving complex learning, planning, and decision-making 
problems. By the bottom-up approach, the key concerns in Edge Computing are categorized into three layers, i.e., Topology, 
Content, and Service. 

For \textbf{Topology}, we pay close attention to the Orchestration of Edge Sites (OES) and Wireless Networking (WN). In 
this paper, we define an edge site as a micro data center with applications deployed, attached to a Small-cell Base 
Station (SBS). OES studies the deployment and installation of wireless telecom equipment and servers. In recent years, research 
efforts on the management and automation of Unmanned Aerial Vehicles (UAVs) became very popular \cite{UAV-begin} \cite{IoT1} 
\cite{review-add5}. UAVs with a small server and an access point can be regarded as moving edge servers with strong 
maneuverability. Therefore, many works explore scheduling and trajectory planning problems with the minimization of energy 
consumption of UAVs. For example, Chen et al. study the power consumption of UAVs by caching the popular contents under 
predictions, where a conceptor-based echo state network (ESN) algorithm is proposed to learn the mobility pattern of users. 
With the help of this effective machine learning technique, the proposed algorithm greatly outperforms benchmarks in terms 
of transmit power and QoE satisfaction. WN studies Data Acquisition and Network Planning. The former concentrates on the 
fast acquisition from rich but highly distributed data at subscribed edge devices while the latter concentrates on network 
scheduling, operation and management. Fast data acquisition includes multiple access, radio resource allocation, and signal 
encoding/decoding. Network planning studies efficient management with protocols and middleware. In recent years, there 
has been an increasing trend in \textit{intelligent} networking, which involves building an intelligent wireless communication mechanism 
by popular AI technologies. For example, Zhu et al. propose \textit{Learning-driven Communication}, which exploits the coupling 
between communication and learning in edge learning systems \cite{learning_driven_communication}. In addition, Sun et al. study the 
resource management in F-RANs (Fog radio access network) with DRL. In order to minimize long-term system power consumption, an MDP 
is formulated and the DQN technique is utilized to make intelligent decisions on the user's equipment communication modes \cite{review-add4}.

For \textbf{Content}, we place an emphasis on Data Provisioning, Service Provisioning, Service Placement, Service Composition and 
Service Caching. For data and service provisioning, the available resources can be provided by remote cloud datacenters and edge 
servers. In recent years, there exist research efforts on constructing lightweight QoS-aware service-based frameworks 
\cite{service-provision} \cite{service_provision}. The shared resources can also come from mobile devices if a proper incentive 
mechanism is employed. Service placement is an important complement to service provisioning, which studies where and how to deploy 
complex services on possible edge sites. In recent years, many works studied service placement from the perspective of Application 
Service Providers (ASPs). For example, Chen et al. try deploying services under limited budget on basic 
communication and computation infrastructures \cite{multi-bandit_learning}. After that, multi-armed bandit theory, an embranchment of reinforcement learning, was 
adopted to optimize the service placement decision. Service composition studies how to select candidate services for composition 
in terms of energy consumption and QoE of mobile end users \cite{trans1} \cite{trans2} \cite{trans3}. It opens up research opportunities 
where AI technologies can be utilized to generate better service selection schemes. Service caching can also be viewed as a complement 
to service provisioning. It studies how to design a caching pool to store the frequently visited data and services. Service caching 
can also be studied in a cooperative way \cite{service_caching}. It leads to research opportunities where multi-agent learning can be 
utilized to optimize QoE in large-scale edge computing systems.

\textbf{For Service}, we focus on Computation Offloading, User Profile Migration, and Mobility Management. Computation offloading 
studies the load balancing of various computational and communication resources in the manner of edge server selection and frequency 
spectrum allocation. More and more research efforts focus on dynamically managing the radio and computational resources for multi-user 
multi-server edge computing systems, utilizing Lyapunov optimization techniques \cite{mywork2} \cite{mywork3}. In recent years, 
optimizing computation offloading decisions via DQN is popular \cite{offloading1} \cite{offloading2}. It models the computation 
offloading problem as a Markov decision process (MDP) and maximize the long-term utility performance. The utility can be composed of 
the above QoE indicators and evolves according to the iterative Bellman equation. After that, the asymptotically optimal computation 
offloading decisions are achieved based on Deep $Q$-Network. User profile migration studies how to adjust the place of user profiles 
(configuration files, private data, logs, etc) when the mobile users are in constant motion. User profile migration is often associated 
with mobility management \cite{trans4}. In \cite{mywork1}, the proposed JCORM algorithm jointly optimizes computation offloading and 
migration by formulating cooperative networks. It opens research opportunities where more advanced AI technologies can be utilized 
to improve optimality. Many existing research efforts study mobility management from the perspective of statistics and probability 
theory. It has strong interests in realizing mobility management with AI.

\subsection{A Recapitulation of AIE} \label{Sce3.3}
The right of the road-map is \textit{AI on edge}. We name this kind of work AIE (i.e. Artificial Intelligence on Edge) since it studies 
how to carry out the training and inference of AI models on the network edge. By top-down decomposition, we divide the research efforts 
in \textit{AI on edge} into three categories: Model Adaptation, Framework Design and Processor Acceleration. Considering that the research 
efforts in Model Adaptation are based on existing training and inference frameworks, let us introduce Framework Design in the first 
place. 

\subsubsection{Framework Design}
Framework design aims at providing a better training and inference architecture for the edge without modifying the existing AI models. 
Researchers attempt to design new frameworks for both Model Training and Model Inference. 

\textbf{\textit{For Model Training:}}
To the best of our knowledge, for model training, all proposed frameworks are distributed, except those knowledge distillation-based 
ones. The distributed training frameworks can be divided into data splitting and model splitting \cite{network_intelligence}. Data splitting can 
be further divided into master-device splitting, helper-device splitting and device-device splitting. The differences lie where the training 
samples come from and how the global model is assembled and aggregated. Model splitting separates neural networks' layers and deploys them 
on different devices. It highly relies on sophisticated pipelines. Knowledge distillation-based frameworks may or may not be decentralized, 
and they rely on transfer learning technologies \cite{transfer}. Knowledge distillation can enhance the accuracy of shallow student networks. 
It first trains a basic network on a basic dataset. After that, the learned features can be transferred to student networks to be trained 
on their datasets, respectively. The basic network can be trained on cloud or edge server while those student networks can be trained by 
numerous mobile end devices with their private data, respectively. We believe that there exist great avenues to be explored in knowledge 
distillation-based frameworks for model training on the edge. 

The most popular work in model training is \textit{Federated Learning} 
\cite{federated1}. Federated Learning is proposed to preserve privacy when training the DNNs in a distributed manner. Without aggregating 
user private data to a central datacenter, Federated Learning trains a series of local models on multiple clients. After that, a global 
model is optimized by averaging the trained gradients of each client. We are not going to elaborate on Federated Learning thoroughly in this article. For 
more details please refer to \cite{federated1}. For the edge nodes with limited storage and computing resource, it is unrealistic to 
train a comprehensive model on their own. Thus, a more applicable way is distributed training, where coordination between edge nodes 
is necessary. For the communication between edge nodes, the challenge is to optimize the global gradient from the distributed local 
models. No matter what learning algorithms is adopted, Stochastic Gradient Decent (SGD) is necessary for model training. Distributed edge 
nodes use SGD to update their local gradients based on their own dataset, which can be viewed as a mini-batch. After that, they send their 
updated gradients to a central node for global model upgrade. In this process, trade-offs between model performance and communication 
overhead has to be considered. If all edge nodes send their local gradients simultaneously, network congestion might be caused. A better 
approach is to selectively choose local gradients which have relatively large improvements. Under this circumstance, the model performance 
of global model can be guaranteed while the communication overheads are reduced.

\textbf{\textit{For Model Inference:}}
Although model splitting is hard to realize for model training, it is a popular approach for model inference. Model 
splitting/partitioning can be viewed as a \textit{framework} for model inference. Other approaches such as model 
compression, input filtering, early-exit and so on can be viewed as adaptations from existing frameworks, which will be 
introduced in the next paragraph and elaborated on carefully in 
Subsection \ref{Sec5.1}. A typical example on model inference on edge is \cite{model_split}, where a DNN is split into two parts and carried 
out collaboratively. The computation-intensive part is running on the edge server while the other is running on the mobile device. The 
problem lies in  where to split the layers and when to exit the intricate DNN according to the constraint on inference accuracy.

\subsubsection{Model Adaptation}
Model Adaptation makes appropriate improvements based on \textit{existing} training and inference frameworks, usually 
Federated Learning, to make them more applicable to the edge. Federated Learning has the potential to run on the 
edge. However, the vanilla version of Federated Learning has a strong demand for communication efficiency since full 
local models are supposed to be sent back to the central server. Therefore, many researchers exploit more efficient model 
updates and aggregation policies. Many works are devoted to reducing cost and increasing robustness while guaranteeing system performance. 
Methods to realize model adaptation include but not limited to Model Compression, Conditional 
Computation, Algorithm Asynchronization and Thorough Decentralization. Model compression exploits the inherent sparsity 
structure of gradients and weights. Possible approaches include but not limited to Quantization, Dimensional Reduction, 
Pruning, Precision Downgrading, Components Sharing, Cutoff and so on. Those approaches can be realized by methods 
such as Singular Value Decomposition (SVD), Huffman Coding, Principal Component Analysis (PCA) and several others. 
Conditional computation is an alternative way to reduce the amount of calculation by selectively turning off some unimportant 
calculations of DNNs. Possible approaches include but not limited to Components Shutoff, Input Filtering, Early Exit, Results 
Caching and so on. Conditional Computation can be viewed as block-wise dropout \cite{conditional1}. Besides, random gossip 
communication can be utilized to reduce unnecessary calculations and model updates. Algorithm Asynchronization trys 
aggregating local models in an asynchronous way. It is designed for overcoming the inefficient and lengthy synchronous 
steps of model updates in Federated Learning. Thoroughly decentralization removes the central aggregator to avoid any possible 
leakage and address the central server's malfunction. The ways to achieve totally decentralization include but not limited to 
blockchain technologies and game-theoretical approaches.

\subsubsection{Processor Acceleration}
Processor Acceleration focuses on structure optimization of DNNs in that the frequently-used computation-intensive 
multiply-and-accumulate operations can be improved. The approaches to accelerate DNN computation on hardware include 
(1) designing special instruction sets for DNN training and inference, (2) designing highly paralleled computing paradigms, 
(3) moving computation closer to memory (near-data processing), etc. Highly parallelized computing paradigms can be divided 
into temporal and spatial architectures \cite{DNN_survey}. The former architectures such as CPUs and GPUs can be accelerated 
by reducing the number of multiplications and increasing throughput. The latter architectures can be accelerated by increasing 
data reuse with data flows. For example, Lee et al. propose an algorithm to accelerate CNN inference \cite{processor_accelerate}. 
The proposed algorithm converts a set of pre-trained weights into values under given precision. It also puts near-data 
processing into practice with an adaptive implementation of memristor crossbar arrays. In the research area of Edge Computing, a 
lot of works hammer at the co-design of Model Adaptation and Processor Acceleration. Considering that Processor Acceleration is 
mainly investigated by AI researchers, this paper will not delve into it. More details on hardware acceleration for DNN processing can be found in \cite{DNN_survey}.

\section{AI for Edge} \label{Sec4}

\begin{figure}[htbp]
  \centering
  \includegraphics[width=3.5in]{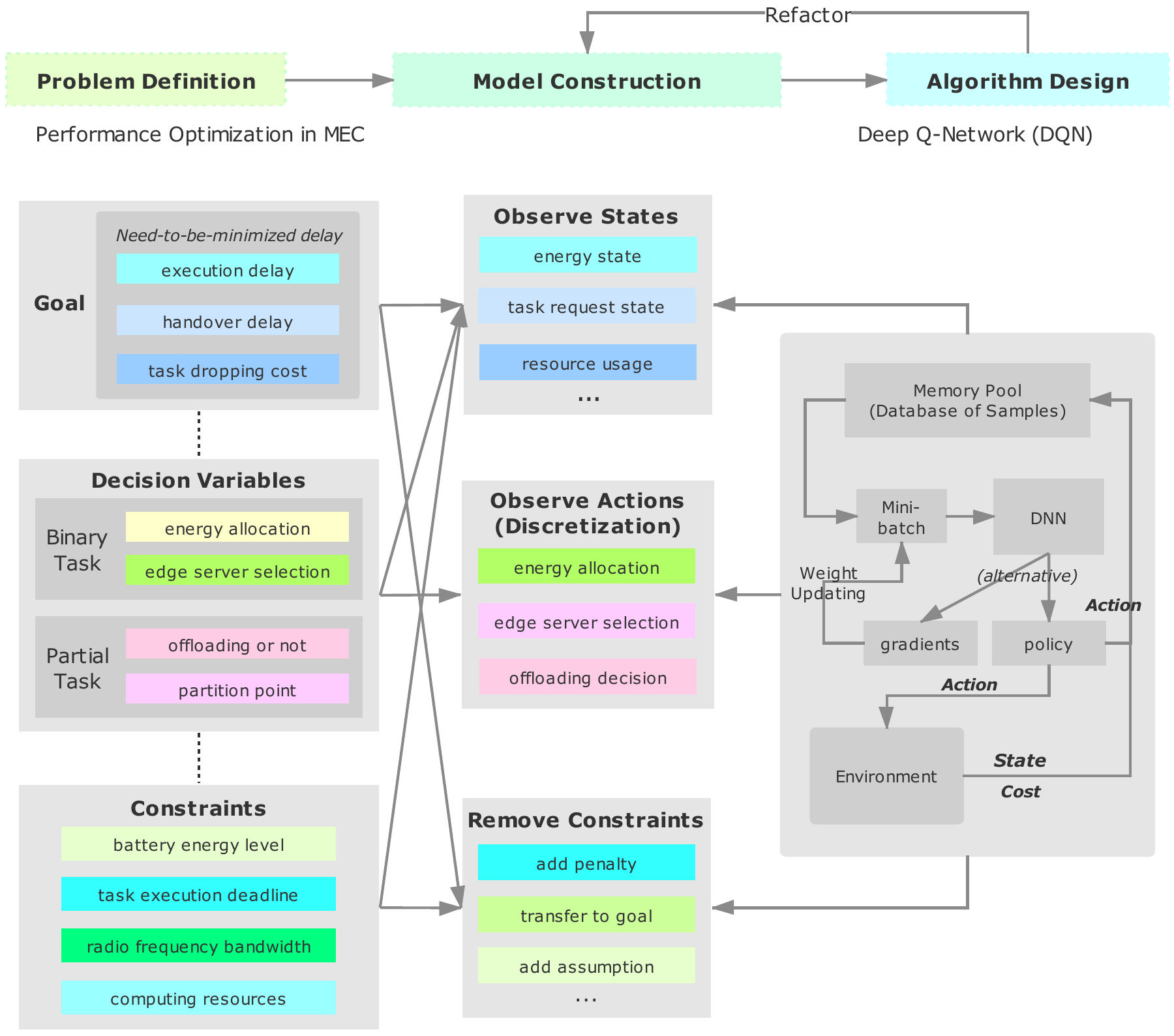}
  \caption{The utilization of AI technology for performance optimization.}
  \label{fig2}
\end{figure}

In Subsection \ref{Sce3.2}, we divide the key issues in Edge Computing into three categories: Topology, Content and Service. 
It just presents a classification and possible research directions but does not provide in-depth analysis on how to apply AI 
technologies to edge to generate more optimal solutions. This Section will remedy this. Fig. \ref{fig2} gives an example of how 
AI technologies are utilized in the Mobile Edge Computing (MEC) environment. Firstly, we need to identify the problem to be 
studied. Take performance optimization as an example, the optimization goal, decision variables, and potential constraints need 
to be confirmed. The need-to-be optimized goal could be the combination of task execution delay, transmission delay and task 
dropping cost. The studied task can be either binary or partial. After that, the mathematical model should be constructed. If the 
long-term stability of system is considered, Lyapunov optimization technique cloud be used to formalize the problem. At last, 
we should design an algorithm to solve the problem. In fact, the model construction is not only decided by the to-be-studied 
problem, but also the to-be-applied optimization algorithms. Take DQN for example, we have to model the problem as an MDP with 
finite \textit{states} and \textit{actions}. Thus, the constraints cannot exist in the long-term optimization problem. The 
most common way is transferring those constraints into penalty and adding the penalty to the optimization goal. 

Considering that current research efforts on \textit{AI for edge} concentrate on Wireless Networking, 
Service Placement, Service Caching and Computation Offloading, we only focus on these topics in the 
following Subsection. For research directions that haven't been explored yet, we are expecting to see more works in due course.

\subsection{State of the Art} \label{Sec4.1}
\subsubsection{Wireless Networking} \label{Sce4.1.1}
The 5G technology promises eMBB, URLLC and mMTC in a real-time and highly dynamic environment. Under the circumstances, researchers reach a 
consensus on that AI technologies should and can be integrated across the wireless infrastructure and mobile users \cite{wireless_survey}. 
We believe that AI should be synergistically applied to achieve intelligent network optimization in a fully online 
manner. One of the typical works in this area is \cite{learning_driven_communication}. This paper advocates a new set of design principles for 
wireless communication on edge with machine learning technologies and models embedded, which are collectively named as 
\textit{Learning-driven Communication}. It can be achieved across the whole process of data acquisition, which are in turn multiple 
access, radio resource management and signal encoding. 

\textbf{Learning-driven multiple access} advocates that the unique characteristics of wireless 
channels should be exploited for functional computation. Over-the-air computation (AirComp) is a 
typical technique used to realize it \cite{over_the_air1} \cite{over_the_air2}. In \cite{broadband} the authors put 
this principle into practice based on Broadband Analog Aggregation (BAA). Concretely, Zhu et al.  
suggest that the simultaneously transmitted model updates in Federated Learning should be analog 
aggregated by exploiting the waveform-superposition property of multi-access 
channels \cite{broadband}. The proposed BAA can dramatically reduce communication latency compared with traditional 
Orthogonal Frequency Division Multiple Access (OFDMA). The work in \cite{over_the_air3} explores the 
\textit{over-the-air computation} for model aggregation in Federated Learning. More specifically, Yang et al. puts the 
principle into practice by modeling the device selection and beamforming design as a sparse and 
low-rank optimization problem, which is computationally intractable \cite{over_the_air3}. To solve the problem 
with a fast convergence rate, this paper proposed a difference-of-convex-functions (DC) representation 
via successive convex relaxation. The numerical results show that the proposed algorithm can achieve 
lower training loss and higher inference accuracy compared with state-of-the-art approaches. 
This contribution can also be categorized as Model Adaptation in \textit{AI on edge}, but it accelerates Federated 
Learning from the perspective of fast data acquisition. 

\textbf{Learning-driven radio resource management} promotes the idea that radio resources should be 
allocated based on the value of transmitted data, not just the efficiency of spectrum utilization. 
Therefore, it can be understood as \textit{importance-aware resource allocation} and an obvious 
approach is \textit{importance-aware retransmission}. In \cite{Retransmission} the authors put the principle into 
practice. This paper proposed a retransmission protocol, named importance-aware automatic-repeat-request 
(importance ARQ). Importance ARQ makes the trade-off between signal-to-noise ratio (SNR) and data 
uncertainty under the desired learning accuracy. It can achieve fast convergence while avoiding 
learning performance degradation caused by channel noise.

\textbf{Learning-driven signal encoding} stipulates that signal encoding should be designed 
by jointly optimizing feature extraction, source coding, and channel encoding. 
A work puts this principle into practice is \cite{Distillation}, which proposes a Hybrid 
Federated Distillation (HFD) scheme based on separate source-channel coding and 
\textit{over-the-air} computing. It adopts sparse binary compression with error accumulation in 
source-channel coding. For both digital and analog implementations over Gaussian multiple-access 
channels, HFD can outperform the vanilla version of Federated Learning in a poor communication environment.
This principle has 
something in common with Dimensional Reduction and Quantization from Model Adaptation in 
\textit{AI on edge}, but it reduces the feature size from the source of data transmission. 
It opens up great research opportunities for the co-design of learning frameworks and data encoding. 

Apart from \textit{Learning-driven Communication}, some works contribute to \textit{AI for 
Wireless Networking} from the perspective of power and energy consumption management. Shen 
et al. utilizes Graph Neural Networks (GNNs) to develop scalable methods for power control in 
$K$-user interference channels \cite{GNN_based}. This paper first models the $K$-user interference 
channel as a complete graph, then it learns the optimal power control with a graph convolutional 
neural network. Temesgene et al. study an energy minimization problem where the baseband 
processes of the virtual small cells powered solely by energy harvesters and batteries can be 
opportunistically executed in a grid-connected edge server \cite{DRL_for_wireless}. Based on multi-agent learning, several 
distributed fuzzy $Q$-learning-based algorithms are tailored. This paper can be viewed as an attempt for coordination with broadcasting. 

As we will expound later, Wireless Networking is often combined with Computation Offloading when it is 
studied in the form of optimization. State of the art of these works is listed in Subsection \ref{Sce4.1.3}.

\subsubsection{Service Placement and Caching} \label{Sce4.1.2}
Many researchers study service placement from the perspective of Application Service Providers (ASPs). 
They model the data and service (it can be compounded and complex) placement problem as a Markov 
Decision Process (MDP) and utilize AI methods such as reinforcement learning to achieve optimal 
placement decision. A typical work implementing this idea is \cite{MAB}. This paper proposes a spatial-temporal algorithm 
based on Multi-armed bandit (MAB) and achieves the optimal placement decisions while learning the benefit. 
Concretely, it studies how many SBSs should be rented for edge service hosting to maximize 
the expected utility up to a finite time horizon. The expected utility is composed of delay reduction 
of all mobile users. After that, a MAB-based algorithm, named SEEN, is proposed to learn the local users’ 
service demand patterns of SBSs. It can achieve the balance between \textit{exploitation} and 
\textit{exploration} automatically according to the fact that whether the set of SBSs is chosen before. 
Another work attempts to integrate AI approaches with service placement is \cite{mywork4}. This 
work jointly decides which SBS to deploy each data block and service component and how much harvested 
energy should be stored in mobile devices with a DQN-based algorithm. This article will not elaborate on DQN. More details can be found in \cite{DRL_survey}.

Service caching can be viewed as a complement to service placement. Edge servers can be equipped with special 
\textit{service cache} to satisfy user demands on popular contents. A wide range of 
optimization problems on service caching are proposed to endow edge servers with learning capability. 
Sadeghi et al. study a sequential fetch-cache decision based on dynamic prices and user 
requests \cite{service_caching}. This paper endows SBSs with efficient fetch-cache decision-making 
schemes operating in dynamic settings. Concretely, it formulates a cost minimization problem with 
service popularity considered. For the long-term stochastic optimization problem, several computationally 
efficient algorithms are developed based on $Q$-learning.

\subsubsection{Computation Offloading} \label{Sce4.1.3}
Computation offloading can be considered as the most active topic when it comes to \textit{AI for edge}.  It studies
the transfer of resource-intensive computational tasks from resource-limited mobile devices to 
edge or cloud. This process involves the allocation of many resources, ranging from CPU cycles 
to channel bandwidth. Therefore, AI technologies with strong optimization abilities have been 
extensively used in recent years. Among all these AI technologies, $Q$-learning and its derivatives, 
DQN, are in the spotlight. For example, Qiu et al. design a $Q$-learning-based algorithm 
for computation offloading \cite{offloading6}. It formulates the computation offloading problem as a 
non-cooperative game in multi-user multi-server edge computing systems and proves that Nash 
Equilibrium exists. Then, this paper proposes a model-free $Q$-learning-based offloading mechanism 
which helps mobile devices learn their long-term offloading strategies to maximize their long-term utilities. 

More works are based on DQN because \textit{the curse of dimensionality} could be overcome with 
non-linear function approximation. For example, Min et al. study the computation 
offloading for IoT devices with energy harvesting in multi-server MEC systems \cite{offloading1}. The 
\textit{need-to-be-maximized} utility formed from overall data sharing gains, task dropping 
penalty, energy consumption and computation delay, which is updated according to the Bellman 
equation. After that, DQN is used to generate the optimal offloading scheme. In \cite{offloading2} 
\cite{offloading3}, the computation offloading problem is formulated as a MDP with \textit{finite} 
states and actions. The state set is composed of the channel qualities, the energy queue, and the 
task queue while the action set is composed of offloading decisions in different time slots. 
Then, a DQN-based algorithm is proposed to minimize the long-term cost. Based on DQN, 
task offloading decisions and wireless resource allocation are jointly optimized to maximize the data 
acquisition and analysis capability of the network \cite{offloading4} \cite{offloading5}. 
The work in \cite{offloading7} studies the knowledge-driven service offloading problem for Vehicle of Internet. 
The problem is also formulated as a long-term planning optimization problem and solved based on DQN. 
In summary, computation offloading problems in various industrial scenarios have been extensively 
studied from all sort of perspectives.

There also exist works who explore the task offloading problem with other AI technologies. For 
example, \cite{offloading8} proposes a long-short-term memory (LSTM) network to predict the task 
popularity and then formulates a joint optimization of the task offloading decisions, computation 
resource allocation and caching decisions. After that, a Bayesian learning automata-based multi-agent 
learning algorithm is proposed for optimality.

\subsection{Grand Challenges} \label{Sec4.2}
Although it is popular to apply AI methods to edge for the generation of better solutions, however, there 
have been many challenges. In the next several Subsections, we list grand challenges across the 
whole theme of \textit{AI for edge} research. These challenges are closely related but each has its own emphasis.

\subsubsection{Model Establishment} \label{Sec4.2.1}
If we want to use AI methods, the mathematical models have to be limited and the formulated 
optimization problem need to be restricted. On one hand, this is because the optimization basis of 
AI technologies, SGD (Stochastic Gradient Descent) and MBGD (Mini-Batch Gradient Descent) methods, may 
not work well if the original search space is constrained. On 
the other hand, especially for MDPs, the state set and action set can not be infinite, and discretization 
is necessary to avoid the curse of dimensionality before further processing. The common solution is changing 
the constraints into a penalty and incorporating them into the global optimization goal. The status quo greatly 
restricts the establishment of mathematical models which leads to performance degradation. It can be viewed 
as a compromise for the utilization of AI methods. Therefore, how to establish an appropriate system model poses great challenges.

\subsubsection{Algorithm Deployment} \label{Sec4.2.2}
The state-of-the-art  works  often  formulate  a  combinatorial and NP-hard optimization problem which 
have fairly high computational complexity. Very  few  works  can achieve an analytic approximate optimal 
solution with convex optimization methods. Actually, for \textit{AI for edge}, the solution mostly comes from 
iterative learning-based  approaches. There are many challenges that face  when these methods are deployed on 
the edge in an online manner. Besides, another ignored challenge is which edge device should undertake the 
responsibility for deploying and running the proposed complicated algorithms. The existing research efforts 
usually concentrate on their specific problems and do not provide the details on that. 

\subsubsection{Balance between Optimality and Efficiency} \label{Sec4.2.3}
Although AI technologies can indeed provide solutions that are optimal, the trade-off between optimality and 
efficiency can not be ignored when it comes to the resource-constrained edge. Thus, how to improve the usability 
and efficiency of edge computing systems for different application scenarios  with AI technologies embedded 
is a severe challenge. The trade-off between optimality and efficiency should be realized based on the 
characteristics of dynamically changing requirements on QoE and the network resource structure. Therefore, 
it is coupling with the service subscribers' pursuing superiority and the utilization of available resources.

\section{AI on Edge} \label{Sec5}
In Subsection \ref{Sce3.3}, we divide the research efforts for \textit{AI on edge} into Model Adaptation, 
Framework Design and Processor Acceleration. The existing frameworks for model training and inference 
are rare. The training frameworks include Federated Learning and Knowledge Distillation while the 
inference frameworks include Model Spitting and Model Partitioning. AI models on edge are by far limited when 
compared to cloud-based predictions because of the relatively limited compute and storage abilities. How to 
carry out the model training and inference on resource-scarce devices is a serious issue. As a result, 
compared with designing new frameworks, researchers in Edge Computing are more interested in improving 
existing frameworks to make them more appropriate for the edge, usually reducing resource occupation. As a 
result, Model Adaptation based on Federated Learning is prosperously developed. As we have mentioned earlier, 
Processor Acceleration will not be elaborated in details. Therefore, we only focus on Model 
Adaptation in the following Subsection. Tab. \ref{table2} 
lists the methods and the correlated papers. Their contributions are also highlighted. 

\begin{table*}[!ht]
  \renewcommand{\arraystretch}{1.2}
  \caption{Methods and the Corresponding Papers.}
  \label{table2}
  \centering
  \begin{tabular}{cc}
    \hline\hline
    \textbf{Methods} & \textbf{Related Papers} \\
    \hline
      Model Compression & 
      \multicolumn{1}{l}{
        \begin{minipage}{3.5in}
          \vskip 4pt
          \begin{itemize}
            \item Sketched updates \& structured updates \cite{compression1}
            \item Communication-efficient secure aggregation \cite{compression2}
            \item Mixed low-bitwidth compression \cite{compression11}
            \item Retraining-after-pruning \cite{compression3}
            \item Compressed RNN (based on Hybrid Matrix Decomposition) \cite{compression5}
            \item Binary Neural Networks (BNNs) \cite{compression4} \cite{compression9} \cite{PCA}
            \item ProNN (based on Stochastic Neighborhood Compression) \cite{compression6}
          \end{itemize}
          \vskip 4pt
        \end{minipage}
      } \\
      \hline
      Conditional Computation & 
      \multicolumn{1}{l}{
        \begin{minipage}{3.5in}
          \vskip 4pt
          \begin{itemize}
            \item Runtime-throttleable block-level gating \cite{conditional2}
          \end{itemize}
          \vskip 4pt
        \end{minipage}
      } \\
      \hline
      Algorithm Asynchronization & 
      \multicolumn{1}{l}{
        \begin{minipage}{3.5in}
          \vskip 4pt
          \begin{itemize}
            \item GoSGD (based on Random-gossip communication) \cite{GoSGD}
            \item GossipGraD (based on Random-gossip communication) \cite{GossipGraD}
          \end{itemize}
          \vskip 4pt
        \end{minipage}
      } \\
      \hline
      Thoroughly Decentralization & 
      \multicolumn{1}{l}{
        \begin{minipage}{3.5in}
          \vskip 4pt
          \begin{itemize}
            \item BlockFL (based on Blockchain) \cite{blockchain_based}
            \item Game-theoretical approach
          \end{itemize}
          \vskip 4pt
        \end{minipage}
      } \\
    \hline
    \hline
  \end{tabular}
\end{table*}

\begin{figure}[htbp]
  \centering
  \includegraphics[width=3.5in]{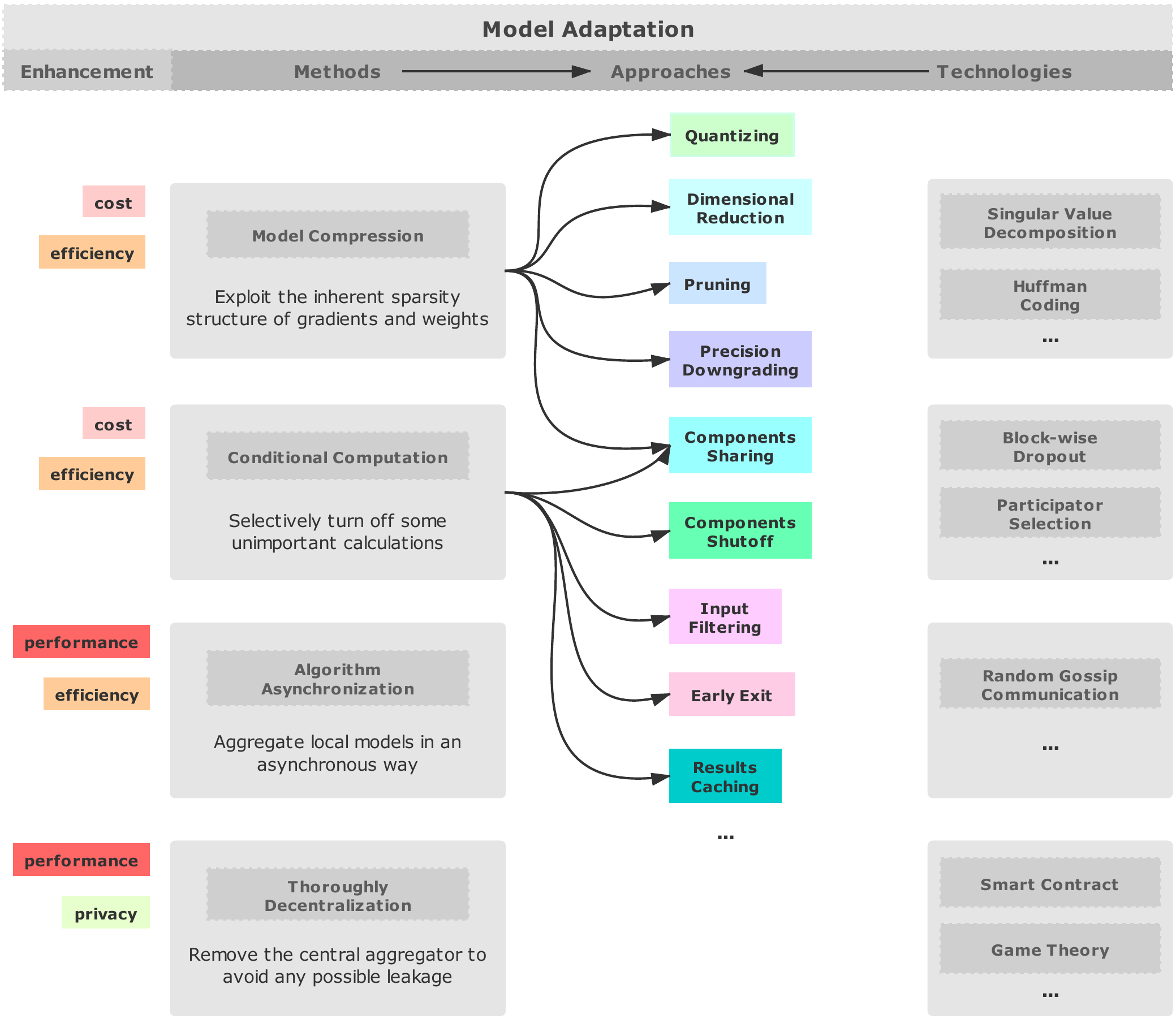}
  \caption{Methods, approaches and technologies of Model Adaptation.}
  \label{fig3}
\end{figure}

\subsection{State of the Art} \label{Sec5.1}
\subsubsection{Model Compression} \label{Sce5.1.1}
As demonstrated in Fig. \ref{fig3}, the approaches for Model Compression include Quantization, 
Dimensionality Reduction, Pruning, Components Sharing, Precision Downgrading and so on. They 
exploit the inherent sparsity structure of gradients and weights to reduce the memory and 
channel occupation as much as possible. The technologies to compress and quantize weights include 
but not limited to Singular Value Decomposition (SVD), Huffman coding and Principal Component Analysis. 
This paper will not provide a thorough introduction to these due to limited space. Considering 
that many works simultaneously utilize the approaches mentioned above, we do not further divide the 
state of the art in Model Compression. One more thing should be clearly noted is that Model 
Compression is suitable for both Model Training and Model Inference. Thus, we do not deliberately distinguish between them.

As we have mentioned earlier, communication efficiency is of the utmost importance for Federated 
Learning. Minimizing the number of rounds of communication is the principal goal when we 
move Federated Learning to the edge because updating the global model might not be achieved if one or more 
local devices are offline or the network is congested. Therefore, a lot of works focus at reducing the 
communication overhead for Federated Learning from various perspectives. Compressing the trained models 
without reducing the inference accuracy is one of the best ways to realize it. 
For example, in \cite{compression1}, \textit{structured updates} and 
\textit{sketched updates} are proposed for reducing the uplink communication costs. For 
\textit{structured updates}, the local update is learnt from a restricted lower-dimensional 
space; for \textit{sketched updates}, the uploading model is compressed before sending to the central 
server. In \cite{compression2}, the authors design a communication-efficient secure aggregation protocol for 
high-dimensional data. The protocol can tolerate up to 33.3\% of participating devices failing to 
complete the protocol, i.e., the system is robust. 
The work in \cite{compression3} suggests that DNNs are typically over-parameterized and their weights have 
significant redundancy. Meanwhile, pruning compensates for the loss in performance. Thus, this paper 
proposes a \textit{retraining-after-pruning} scheme. It retrains the DNN on new data while the 
pruned weights stay constant. The scheme can reduce the resource occupation while guaranteeing learning accuracy. The work in 
\cite{compression11} exploits mixed low-bitwidth compression. It works 
on determining the minimum bit precision of each activation and weight under the given constraints 
on memory. The authors in \cite{compression4} use Binarized Neural Networks (BNNs), which have binary weights and 
activations to replace regular DNNs. This is a typical exploration of quantization. Analogously, 
Chakraborty et al. propose hybrid network architectures combing binary and full-precision sections 
to achieve significant energy efficiency and memory compression with performance guaranteed \cite{compression9}. 
Thakker et al. study a compressed RNN cell implementation called Hybrid Matrix 
Decomposition (HMD) for model inference \cite{compression5}. It divides the matrix of 
network weights into two parts: an unconstrained upper half and a lower half composed 
of rank-1 blocks. The output features are composed of the rich part (upper) and the barren 
part (lower). This is an imaginative variation on compression, compared with traditional 
pruning or quantization. The numerical results show that it can not only achieve a faster run-time than 
pruning and but also retain more model accuracy than matrix factorization.

Some works also explore model compression based on partitioned DNNs. For example, Li et al. proposes 
an auto-tuning neural network quantization framework for collaborative inference between edge and cloud \cite{compression8}. 
Firstly, DNN is partitioned. The first part is quantized and executed on the edge devices while the 
second part is executed in cloud \textit{with full-precision}. The work in \cite{compression10} proposes a framework to 
accelerate and compress model training and inference. It partitions DNNs into multiple sections according 
to their depth and constructs classifiers upon the intermediate features of different sections. Besides, 
the accuracy of classifiers is enhanced by knowledge distillation. 

Apart from Federated Learning, there exist works that probe into the execution of statistical 
learning models or other popular deep models such as ResNet and VGG architectures on 
resource-limited end devices. For example, Gupta et al. propose ProtoNN, 
a compressed and accurate $k$-Nearest Neighbor (kNN) algorithm \cite{compression6}. ProtoNN learns a small number 
of prototypes to represent the entire training set by Stochastic Neighborhood Compression (SNC) 
\cite{compression7}, and then projects the entire data in a lower dimension with a sparse 
projection matrix. It jointly optimizes the projection and prototypes with explicit model size constraint. 
Chakraborty et al. proposes Hybrid-Net which has both binary and high-precision 
layers to reduce the degradation of learning performance \cite{PCA}. Innovatively, this paper 
leverages PCA to identify significant layers in a binary network, other than dimensionality 
reduction. The \textit{significance} here is identified based on the ability of a layer to expand 
into higher dimensional space.

Model Compression is currently a very active direction in \textit{AI on edge} because it is easy to implement. 
However, the state-of-the-art works are usually not tied to specific application scenarios of 
edge computing systems. There are opportunities for new works that construct edge platforms and hardware.

\subsubsection{Conditional Computation} \label{Sec5.1.2}
As demonstrated in Fig. \ref{fig3}, the approaches for Conditional Computation include Components 
Sharing, Components Shutoff, Input Filtering, Early Exit, Results Caching and so on. To put it simply, 
Conditional computation is selectively turning off some unimportant calculations. Thus it can be 
viewed as block-wise dropout \cite{conditional1}. A lot of works devote themselves to ranking and selecting 
the most worthy part for computation or early stop if the confident threshold is achieved. 
For example, Hostetler et al. instantiate a \textit{runtime-throttleable} neural network which can 
adaptively balance learning accuracy and resource occupation in response to a control signal \cite{conditional2}. It 
puts Conditional Computation into practice via block-level gating. 

This idea can also be put into use for participator selection. It selects the most valuable participators 
in Federated Learning for model updates. The valueless participators will not engage the aggregation of the global model. 
To the best of our knowledge, currently, there is no work dedicated to participator selection. We 
are eagerly looking forward to exciting works on it.

\subsubsection{Algorithm Asynchronization} \label{Sec5.1.3}
As demonstrated in Fig. \ref{fig3}, Algorithm Asynchronization attempts to aggregate local models in an asynchronous way for Federated 
Learning. As we have mentioned before, the participating local users have a significant probability 
of failing to complete the model upload and download due to the wireless network congestion. Apart 
from model compression, another way is exchanging weights and gradients \textit{peer-to-peer} to 
reduce the high concurrency on wireless channels. Random-gossip Communication is a typical example. 
Based on randomized gossip algorithms, Blot et al. propose GoSGD to train DNNs asynchronously \cite{GoSGD}. 
The most challenging problem for gossip training is the degradation 
of convergence rate in large-scale edge systems. To overcome the issue, Daily et al. introduce 
GossipGraD, which can reduce the communication complexity greatly to ensure the fast convergence 
\cite{GossipGraD}.

\subsubsection{Thorough Decentralization} \label{Sce5.1.4}
As demonstrated in Fig. \ref{fig3}, Thorough Decentralization attempts to remove the central 
aggregator to avoid any possible leakage. Although Federated Learning does not require consumers’ 
private data, the model updates still contain private information as some trust of the server 
coordinating the training is still required. To avoid privacy leaks altogether, blockchain 
technology and game-theoretical approaches can assist in total decentralization.

By leveraging blockchain, especially smart contracts, the central server for model aggregating 
is not needed anymore. As a result, collapse triggered by model aggregation can be avoided. 
Besides, user privacy can be protected. We believe that the blockchain-based Federated 
Learning will become a hot field and prosperous direction in the coming years. There exists works that put 
it into practice. In \cite{blockchain_based}, the proposed blockchain-based federated 
learning architecture, BlockFL, takes edge nodes as miners. Miners exchange and verify all the 
local model updates contributed by each device and then run the Proof-of-Work (PoW). The miner
who firstly completes the PoW generates a new block and receives the mining reward from the 
blockchain network. At last, each device updates its local model from the freshest block. In this 
paper, blockchain is effectively integrated with Federated Learning to build a trustworthy edge 
learning environment.

\subsection{Grand Challenges} \label{Sec5.2}
The grand challenges for AI on edge are listed from the perspective of data availability, model selection, and 
coordination mechanism, respectively. 

\subsubsection{Data Availability}
The toughest challenge lies in the availability and usability of raw training data because usable data is 
the beginning of everything. Firstly, a proper incentive mechanism may be necessary for data provisioning 
from mobile users. Otherwise, the raw data may not be available for model training and inference. 
Besides, the raw data from various end devices could have an obvious bias, which can greatly affect the 
learning performance. Although Federated Learning can overcome the problem caused by non-i.i.d. 
samples to a certain extent, the training procedure still faces great difficulties on the design 
of robust communication protocol. Therefore, there are huge challenges in terms of data availability.

\subsubsection{Model Selection}
At present, the selection of \textit{need-to-be-trained} AI models faces severe challenges in the following 
aspects, across from the models themselves to the training frameworks and hardware. Firstly, how to select 
the befitting threshold of learning accuracy and scale of AI models for quick deployment and delivery. 
Secondly, how to select probe training frameworks and accelerator architectures under the 
limited resources. Model selection is coupling with resource allocation and management, thus the problem 
is complicated and challenging.

\subsubsection{Coordination Mechanism}
The proposed methods on Model Adaptation may not be pervasively serviceable because there could 
be a huge difference in computing power and communication resources between heterogeneous edge devices. 
It may lead to that the same method achieves different learning results for different clusters of 
mobile devices. Therefore, the compatibility and coordination between heterogeneous edge devices are of 
great essence. A flexible coordination mechanism between cloud, edge, and device in both hardware and 
middleware is imperative and urgently needed to be designed. It opens up research opportunities on 
a uniform API interface on edge learning for ubiquitous edge devices.

\section{Concluding Remarks} \label{Sec6}
Edge Intelligence, although still in its early stages, has attracted more and more researchers 
and companies to get involved in studying and using it. This article attempts to provide possible research 
opportunities through a succinct and effective classification. Concretely, we first discuss the relation 
between Edge Computing and Artificial Intelligence. We believe that they promote and reinforce each other. 
After that, we divide Edge Intelligence into \textit{AI for edge} and \textit{AI on edge} and sketch 
the research road-map. The former focuses on providing a better solution to the key concerns in Edge 
Computing with the help of popular and rich AI technologies while the latter studies how to carry 
out the training and inference of AI models, on edge. Either \textit{AI for edge} or \textit{AI on edge}, 
the research road-map is presented in a hierarchical architecture. By the bottom-up approach, we divide 
research efforts in Edge Computing into Topology, Content, and Service and introduce some examples on how to 
energize edge with intelligence. By top-down decomposition, we divide the research efforts in \textit{AI on edge} 
into Model Adaptation, Framework Design, and Processor Acceleration and introduce some existing research 
results. Finally, we present the state of the art and grand challenges in several hot topics for both 
\textit{AI for edge} and \textit{AI on edge}. We attempted to provide some enlightening thoughts on the 
emerging field of Edge Intelligence. We hope that this paper can stimulate fruitful discussions on potential 
future research directions for Edge Intelligence.

\ifCLASSOPTIONcaptionsoff
  \newpage
\fi



\bibliographystyle{IEEEtran}
\bibliography{IEEEabrv,ref}
%



%
\begin{IEEEbiography}[{\includegraphics[width=1in,height=1.25in,clip,keepaspectratio]{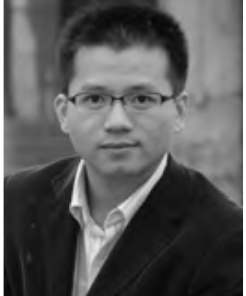}}]{Shuiguang Deng}
is currently a full professor at the First Affiliated Hospital of Zhejiang University School of Medicineat as well 
as the College of Computer Science and Technology in Zhejiang University, China. He previously worked at the 
Massachusetts Institute of Technology in 2014 and Stanford University in 2015 as a visiting scholar. His research 
interests include Edge Computing, Service Computing, Mobile Computing, and Business Process Management. He 
serves as the associate editor for the journal IEEE Access and IET Cyber-Physical Systems: Theory \& Applications. 
Up to now, he has published more than 100 papers in journals and refereed conferences. In 2018, he was granted the 
Rising Star Award by IEEE TCSVC. He is a fellow of IET and a senior member of IEEE.
\end{IEEEbiography}

\begin{IEEEbiography}[{\includegraphics[width=1in,height=1.25in,clip,keepaspectratio]{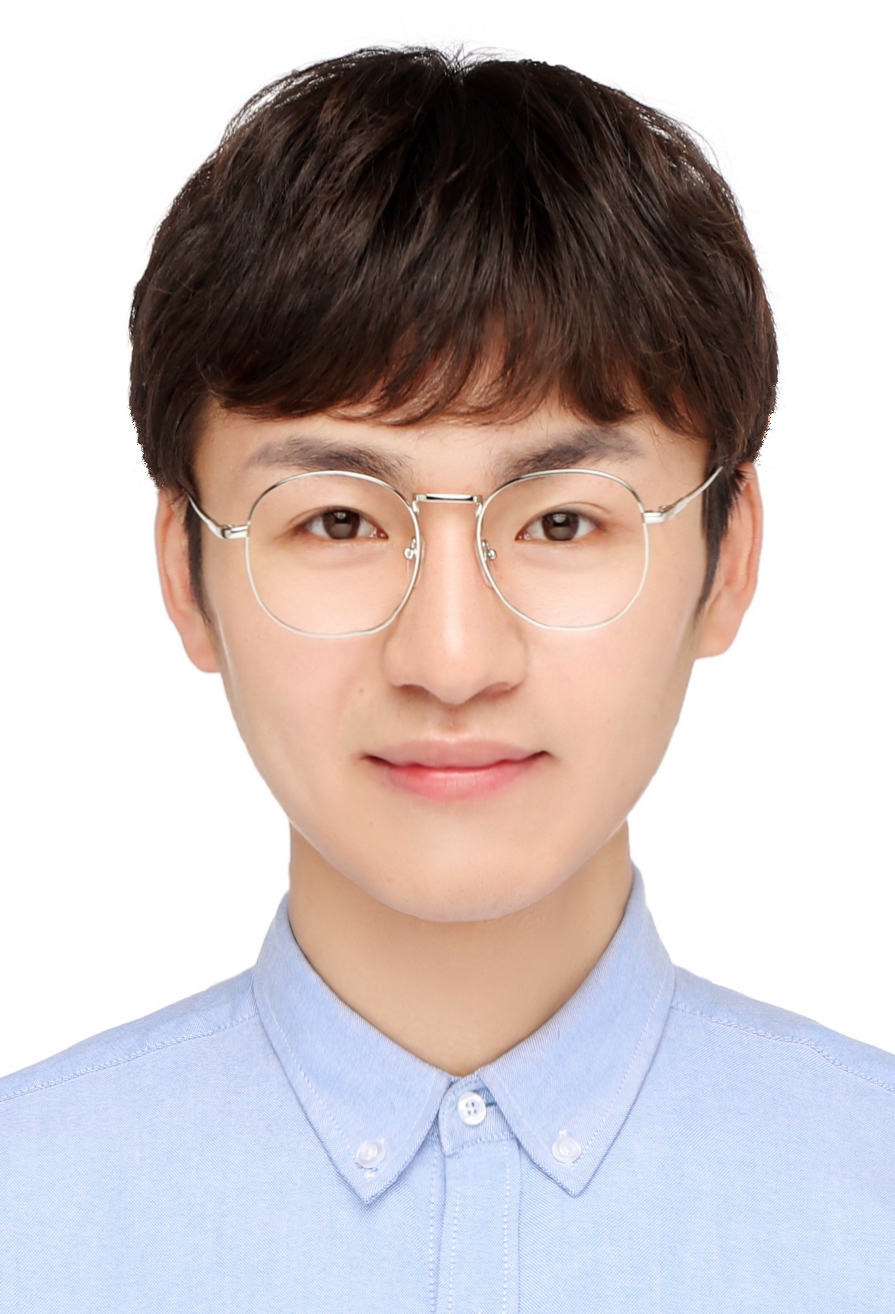}}]{Hailiang Zhao}
  received the B.S. degree in 2019 from the school of computer science 
  and technology, Wuhan University of Technology, Wuhan, China. He is currently 
  pursuing the Ph.D. degree with the College of Computer Science and Technology, 
  Zhejiang University, Hangzhou, China. He has been a recipient of the Best Student 
  Paper Award of IEEE ICWS 2019. His research interests include edge computing, 
  service computing and machine learning.
\end{IEEEbiography}

\begin{IEEEbiography}[{\includegraphics[width=1in,height=1.25in,clip,keepaspectratio]{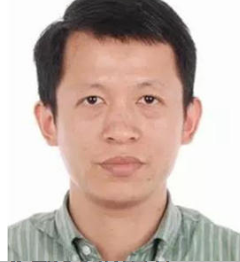}}]{Weijia Fang}
  received Master’s degree in oncology and Doctorate in oncology/surgery in 2005 and 2013 from Zhejiang University, 
  China. He works in the department of Medical Oncology, the First Affiliated Hospital, Zhejiang University School of 
  Medicine, China. He has been author and co-author of several original research publications in national and 
  international peer-reviewed scientific and medical journals.
\end{IEEEbiography}

\begin{IEEEbiography}[{\includegraphics[width=1in,height=1.25in,clip,keepaspectratio]{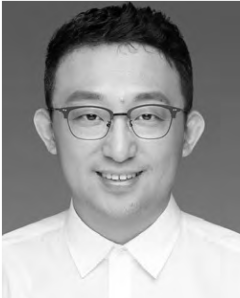}}]{Jianwei Yin}
  received the Ph.D. degree in computer science from Zhejiang University (ZJU) in 2001. He was a Visiting Scholar 
  with the Georgia Institute of Technology. He is currently a Full Professor with the College of Computer Science, 
  ZJU. Up to now, he has published morethan 100 papers in top international journals and conferences. His current 
  research interests include service computing and business process management. He is an Associate Editor of the 
  IEEE Transactions on Services Computing.
\end{IEEEbiography}

\begin{IEEEbiography}[{\includegraphics[width=1in,height=1.25in,clip,keepaspectratio]{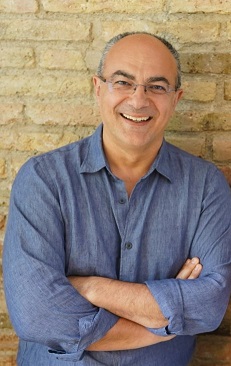}}]{Schahram Dustdar}
  is a Full Professor of Computer Science (Informatics) with a focus on Internet Technologies heading the Distributed 
  Systems Group at the TU Wien. He is Chairman of the Informatics Section of the Academia Europaea (since December 9, 2016). 
  He is elevated to IEEE Fellow (since January 2016). From 2004-2010 he was Honorary Professor of Information Systems 
  at the Department of Computing Science at the University of Groningen (RuG), The Netherlands.

  From December 2016 until January 2017 he was a Visiting Professor at the University of Sevilla, Spain and from January 
  until June 2017 he was a Visiting Professor at UC Berkeley, USA. He is a member of the IEEE Conference Activities Committee 
  (CAC) (since 2016), of the Section Committee of Informatics of the Academia Europaea (since 2015), a member of the Academia 
  Europaea: The Academy of Europe, Informatics Section (since 2013). He is recipient of the ACM Distinguished Scientist award 
  (2009) and the IBM Faculty Award (2012). He is an Associate Editor of IEEE Transactions on Services Computing, ACM Transactions 
  on the Web, and ACM Transactions on Internet Technology and on the editorial board of IEEE Internet Computing. He is the 
  Editor-in-Chief of Computing (an SCI-ranked journal of Springer).
\end{IEEEbiography}

\begin{IEEEbiography}[{\includegraphics[width=1in,height=1.25in,clip,keepaspectratio]{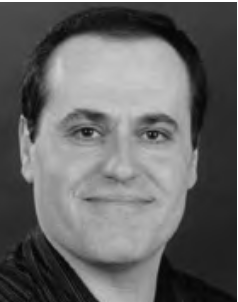}}]{Albert Y. Zomaya}
  is the Chair Professor of High Performance Computing \& Networking in the School of Computer Science, University of 
  Sydney, and he also serves as the Director of the Centre for Distributed and High Performance Computing. Professor Zomaya 
  published more than 600 scientific papers and articles and is author, co-author or editor of more than 30 books. He is the 
  Editor in Chief of the IEEE Transactions on Sustainable Computing and ACM Computing Surveys and serves as an associate 
  editor for several leading journals. Professor Zomaya served as an Editor in Chief for the IEEE Transactions on Computers 
  (2011-2014). He is a Chartered Engineer, a Fellow of AAAS, IEEE, and IET. Professor Zomaya’s research interests are in the 
  areas of parallel and distributed computing and complex systems.
\end{IEEEbiography}






\end{document}